\documentclass[fleqn,usenatbib]{rasti}

\usepackage{newtxtext,newtxmath}

\usepackage[T1]{fontenc}

\DeclareRobustCommand{\VAN}[3]{#2}
\let\VANthebibliography\thebibliography
\def\thebibliography{\DeclareRobustCommand{\VAN}[3]{##3}\VANthebibliography}

\usepackage{graphicx}	
\usepackage{float}
\graphicspath{Figures/}
\usepackage[color=blue!20!white, obeyFinal]{todonotes}

\title[Generative imaging for radio interferometry]{Generative imaging for radio interferometry with fast uncertainty quantification
}

\author[M. Mars et al.]{
Matthijs Mars,$^{1,2}$\thanks{E-mail: academic@matthijsmars.com}
Tobías I.~Liaudat,$^{3}$
Jessica J.~Whitney,$^{1}$
Marta M.~Betcke,$^{4}$
Jason D.~McEwen$^{1,5}$\thanks{E-mail: jason.mcewen@ucl.ac.uk}
\\
$^{1}$Mullard Space Science Laboratory (MSSL), University College London (UCL), Dorking RH5 6NT, UK\\
$^{2}$Leiden Observatory, Leiden University, Leiden 2333 CC, The Netherlands\\
$^{3}$ IRFU, CEA, Université Paris-Saclay, F-91191 Gif-sur-Yvette, France \\
$^{4}$Department of Computer Science, University College London (UCL), London WC1E 6BT, UK \\
$^{5}$Alan Turing Institute, London NW1 2DB, UK
}

\date{Accepted XXX. Received YYY; in original form ZZZ}

\pubyear{2025}

\usepackage{soul}

\hypersetup{
  colorlinks=true,
}

\begin{document}
\label{firstpage}
\pagerange{\pageref{firstpage}--\pageref{lastpage}}
\maketitle

\begin{abstract}
  With the rise of large radio interferometric telescopes, particularly the SKA, there is a growing demand for computationally efficient image reconstruction techniques. Existing reconstruction methods, such as the CLEAN algorithm or proximal optimisation approaches, are iterative in nature, necessitating a large amount of compute. These methods either provide no uncertainty quantification or require large computational overhead to do so. Learned reconstruction methods have shown promise in providing efficient and high quality reconstruction. 
In this article we explore the use of generative neural networks that enable efficient approximate sampling of the posterior distribution for high quality reconstructions with uncertainty quantification. Our RI-GAN framework, builds on the regularised conditional generative adversarial network (rcGAN) framework by integrating a gradient U-Net (GU-Net) architecture - a hybrid reconstruction model that embeds the measurement operator directly into the network. This framework uses Wasserstein GANs to improve training stability in combination with regularisation terms that combat mode collapse, which are typical problems for conditional GANs. This approach takes as input the dirty image and the point spread function (PSF) of the observation and provides efficient, high-quality image reconstructions that are robust to varying visibility coverages, generalises to images with an increased dynamic range, and provides informative uncertainty quantification. Our methods provide a significant step toward computationally efficient, scalable, and uncertainty-aware imaging for next-generation radio telescopes.
\end{abstract}


\section{Introduction}
In order to make reliable scientific conclusions from astronomical observations, it is crucial to quantify the uncertainties in the image reconstructions. 
However, most current radio interferometric imaging methods do not provide uncertainty quantification, and methods that do are often computationally expensive and not scalable to the large data volumes expected from next-generation radio telescopes.
Additionally, the advent of large radio interferometer telescopes such as the Square Kilometre Array \citep[SKA,][]{dewdneySquareKilometreArray2009} leads to an increasing need for efficient image reconstruction methods, making the provision of uncertainty quantification significantly more challenging. 
One of the main challenges in radio interferometric imaging is dealing with the increasingly large amount of visibilities acquired by the telescopes as this increases the computational complexity of evaluating the measurement operator of the telescope. 

Current reconstruction methods, such as CLEAN \citep{hogbomApertureSynthesisNonregular1974,bhatnagarScaleSensitiveDeconvolution2004,bhatnagarCorrectingDirectiondependentGains2008,stewartMultiplebeamCLEANImaging2011,offringaWSCleanImplementationFast2014}, as well as more recently developed methods based on compressed sensing techniques \citep[e.g.][]{carrilloSparsityAveragingReweighted2012,carrilloSparsityAveragingCompressive2013, pratleyRobustSparseImage2018} are based on iterative optimisation. 
These methods are computationally expensive as they have to evaluate the measurement operator at each iteration and typically provide only a point estimate without quantifying the uncertainties. 
While scalable and highly distributed methods have been developed to manage large data volumes \citep[e.g.][]{pratleyDistributedParallelSparse2019} as well as some methods to estimate the uncertainties in the reconstructions \citep{caiUncertaintyQuantificationRadio2018, liaudatScalableBayesianUncertainty2024,cherifUncertaintyQuantificationFast2024}, these methods still struggle with the data volumes expected from the SKA.

Statistical methods like Markov chain Monte Carlo (MCMC) are capable of performing Bayesian inference by sampling the full posterior distribution and quantify the uncertainties in the reconstruction \citep[e.g.][]{sutterProbabilisticImageReconstruction2014,caiUncertaintyQuantificationRadio2018a}. 
However, these methods are particularly computationally expensive due to the iterative nature of MCMC sampling techniques, where the measurement operator needs to be evaluated a large amount of times. Therefore, these methods do not scale well to larger data volumes.

An alternative Bayesian approach, that does not require sampling of the full posterior distribution, is the \texttt{RESOLVE} method \citep{junklewitzRESOLVENewAlgorithm2016,greinerFastRESOLVEFastBayesian2017,arrasRadioImagingInformation2018a}. This method uses a variational approach to approximate the posterior distribution using a technique called metric Gaussian variational inference \citep[MGVI,][]{knollmullerMetricGaussianVariational2020} and uses samples from this approximate posterior to estimate the uncertainties in the reconstruction.

Recent advances in machine learning for radio interferometric image reconstruction have shown promise in terms of superior reconstruction quality and computational efficiency. 
Learned regularisation methods, which employ a learned prior or regulariser in a variational regularisation framework, have demonstrated high reconstruction quality, robustness to varying measurement operators, and good generalising to unseen data \citep[e.g.][]{terrisImageReconstructionAlgorithms2022,terrisPlugandplayImagingModel2023,liaudatScalableBayesianUncertainty2024}. 
Post-processing methods, which are computationally more efficient than iterative based approaches, remove artifacts from an initial reconstruction and provide high-quality results \citep[e.g.][]{allamjrRadioInterferometricImage2016,connorDeepRadiointerferometricImaging2022,marsLearnedInterferometricImaging2023b,marsLearnedRadioInterferometric2025}. 
These methods are less robust to varying visibility coverages and require full retraining for each new observation, though more efficient training strategies can be used to mitigate this somewhat \citep{marsLearnedRadioInterferometric2025}.
Unrolled methods, which are trained to perform a fixed number of learned updates in an iterative algorithm, have shown to provide high quality reconstructions and require far fewer iterations than fully iterative methods \citep[e.g.][]{marsLearnedRadioInterferometric2025,aghabiglouR2D2DeepNeural2024,dabbechCLEANingCygnusDeep2023}. 
These methods explicitly include the measurement operator in the reconstruction network, significantly improving reconstruction quality compared to post-processing methods and providing robustness to varying visibility coverages \citep{marsLearnedRadioInterferometric2025}.

While most learned methods only provide a single point estimate of the image, recent approaches consider the inclusion of uncertainty quantification. 
One strategy is to train multiple versions of the learned reconstruction models to represent different realisations of the prior distribution, which can then estimate the model uncertainty in the reconstruction by providing multiple reconstructions and examining their variance \citep{terrisPlugandplayImagingModel2023}. 
Another strategy uses the theory of probability concentration (extending the work of \citet{caiUncertaintyQuantificationRadio2018} to the learned setting) to provide approximate uncertainty quantification by using a convex learned prior without the need for sampling or generating several realisations \citep{liaudatScalableBayesianUncertainty2024}.

Conditional generative imaging methods create a generating function that maps a latent code vector (typically a vector of random noise) to an image sample of the posterior distribution, conditioned on the observed data. 
Popular generative models in astronomical inverse problems have been either score-based \citep[e.g.][]{remyProbabilisticMassMapping2023,diaIRISBayesianApproach2025} or diffusion-based \citep[e.g.][]{wangConditionalDenoisingDiffusion2023a, drozdovaRadioastronomicalImageReconstruction2024} methods. 
These methods, despite providing high quality reconstructions and diverse posterior samples, are orders of magnitude slower in generating samples than methods successfully applied in other imaging problems based on conditional generative adversarial networks \citep[cGANs, e.g.,][]{isolaImageToImageTranslationConditional2017,zhaoSynthesizingRetinalNeuronal2018, zhao2021large, AdlerÖktem+2025+359+412}, conditional variational autoencoders \citep[cVAEs, e.g., ][]{edupugantiUncertaintyQuantificationDeep2021,tonoliniVariationalInferenceComputational2020,sohnLearningStructuredOutput2015}, or conditional normalizing flows \citep[cNFs, e.g.,][]{ardizzoneGuidedImageGeneration2019,winklerLearningLikelihoodsConditional2023,sunDeepProbabilisticImaging2021}.

In this work we focus on using cGANs for radio interferometric image reconstruction with approximate posterior samples to build our radio interferometry GAN (RI-GAN) framework. While cGANs have been known to suffer from mode-collapse, resulting in reduced sample diversity, recent work by \citet{bendelRegularizedConditionalGAN2023} has introduced a regularised cGAN (rcGAN). This approach uses regularisation to prevent mode-collapse and promote sample diversity, with successful applications in inpainting and magnetic resonance imaging (MRI) reconstruction problems. Extensions to astronomical imaging, e.g., by \citet[][]{whitneyUsingConditionalGANs2024,whitneyGenerativeModellingMassmapping2024} in mass mapping, show promise in providing both efficient and diverse posterior samples.

A significant challenge in constructing learned methods for radio interferometric imaging is the variability of the measurement operator, which depends on the telescope's \emph{uv}-coverage and varies for each observation based on the telescope's pointing, the duration of the observation, and the rotation of the Earth. 
In order to encode this information into the reconstruction, we additionally condition the generator of the rcGAN with the observation's point spread function (PSF). 
Previous work has shown that conditioning learned reconstruction methods on the measurement operator through explicit embedding of the operator in the reconstruction network significantly enhances reconstruction quality, robustness to varying visibility coverages, and generalisation to out-of-distribution data \citep{marsLearnedRadioInterferometric2025}. 
Building on this, we also present an rcGAN generator that explicitly embeds the operator within the network. This additional conditioning enhances the reconstruction, robustness, and generalisation capabilities of our method.

The remainder of this article is structured as follows. Section~\ref{sec:radio-imaging} provides an overview of the radio interferometric imaging problem, its challenges, and a discussion of established reconstruction methods. 
In Section~\ref{sec:generative-imaging} we introduce the concept of generative imaging and discuss the rcGAN framework. 
Section~\ref{sec:proposed-methods} describes our two proposed RI-GAN models, one based on a U-Net and the other based on a gradient U-Net \citep[GU-Net;][]{marsLearnedInterferometricImaging2023b,marsLearnedRadioInterferometric2025} generator. 
It also details the implementation of the rcGAN for radio interferometric imaging and the training procedure. 
In Section~\ref{sec:experiments} we evaluate the performance of the rcGAN on simulated data and compare the two models. 
We also investigate the number of samples needed for the reconstructions and uncertainty estimates, and the correlation of the uncertainties with the errors of the reconstructions.
Additionally, we evaluate the models on simulated measurements from the 30 Doradus region. 
Finally, in Section~\ref{sec:conclusion} we summarise our findings and discuss future work.
\section{Radio interferometric imaging}\label{sec:radio-imaging}
Aperture synthesis in radio interferometric imaging is an essential technique in radio astronomy to obtain high-resolution images of the radio sky. 
The process involves measuring visibilities, which are Fourier components of the sky brightness distribution, using a an array of radio telescopes. 
However, reconstructing high-fidelity images from these visibilities is a challenging problem due to the non-uniform sampling of the Fourier domain, the presence of noise, and the wide-field effects that occur when the baselines are not co-planar.

The measurement process for a radio telescope can be described as \citep[see, e.g.,][]{thompsonInterferometrySynthesisRadio2017}
\begin{equation}
  \begin{aligned}
    \mathcal{V} & (u, v, w) = \int_{-\infty}^{\infty} \int_{-\infty}^{\infty} \frac{1}{\sqrt{1-l^2-m^2}} A_N(l, m) I(l, m) \\
                & \times \exp \left\{-i 2 \pi\left[u l+v m+w\left(\sqrt{1-l^2-m^2}-1\right)\right]\right\} d l d m,
  \end{aligned}
\end{equation}
where $\mathcal{V}(u, v, w)$ represents the acquired visibility, $A_N(l, m)$ is the primary beam pattern, and $I(l, m)$ the sky brightness distribution. 
If the telescope array is small, the baselines can be considered to be co-planar and the $w$-term can be ignored, reducing the equation to a 2D Fourier transform. 
This results in the sky brightness distribution convolved with the point spread function (PSF) of the telescope. 
If the baselines are not co-planar, the $w$-term cannot be ignored and needs to be included in the measurement operator and for large fields of view direction dependent effects (DDEs) need to be accounted for \citep{smirnovRevisitingRadioInterferometer2011}.

The measurement acquisition process can be summarised as
\begin{equation}\label{eq:inverse-problem}
    \boldsymbol{y} = \boldsymbol{\Phi} \boldsymbol{x} + \boldsymbol{n},
\end{equation}
where $\boldsymbol{y} \in \mathbb{C}^M$ are the measured visibilities, $\boldsymbol{x} \in \mathbb{R}^N$ is the image to be reconstructed, $\boldsymbol{n} \in \mathbb{C}^M$ the noise in the measurements, and $\boldsymbol{\Phi}: \mathbb{R}^{N} \rightarrow \mathbb{C}^M$ is the measurement operator of the telescope. 
The measurement operator is typically modelled using a non-uniform fast Fourier transform \citep[NUFFT, ][]{duijndamNonuniformFastFourier1997,fesslerNonuniformFastFourier2003}, which grids the visibilities to a uniform grid in order to use the fast Fourier transform \citep[FFT,][]{duttFastFourierTransforms1993} to speed up the computation. 
More details on implementing this measurement operator in the context of radio interferometry can be found in e.g., \citet{pratleyRobustSparseImage2018,marsLearnedInterferometricImaging2023b}.

For the co-planar baseline case, the response of the telescope can be described as a convolution with the PSF of the telescope
\begin{equation}
    \boldsymbol{x}_{\text{dirty}} = \boldsymbol{\Phi}^* \boldsymbol{y} = \boldsymbol{\Phi}^* \boldsymbol{\Phi} \boldsymbol{x}  = \boldsymbol{I}_{\text{PSF}} * \boldsymbol{x},
\end{equation}
where $x_{\text{dirty}} \in \mathbb{R}^N$ is the naturally weighted dirty image,  $\boldsymbol{\Phi}^* : \mathbb{C}^M \rightarrow \mathbb{R}^N$ is the adjoint of the measurement operator and where the PSF of the telescope is described as $\boldsymbol{I}_{\text{PSF}} \triangleq \boldsymbol{\Phi}^* \boldsymbol{\Phi} \ \boldsymbol{\delta}$, with $\boldsymbol{\delta} \in \mathbb{R}^N$ being an image with a value of 1 in the centre pixel and 0 elsewhere. 
However, the wide-field effects and DDEs that occur from large fields of view and non-coplanar baselines are not modelled by this convolution. 

\section{Generative imaging}\label{sec:generative-imaging}
In this work we focus on using cGANs to provide our generative framework. 
Where unconditional GANs map from a latent code vector to an image, conditional GANs are also conditioned on the observed data. 
cGANs consist of two parts: a generator network $\boldsymbol{G}_\theta$ that generates samples from the posterior distribution and a discriminator network $\boldsymbol{D}_\phi$ that tries to distinguish between true samples and generated samples. 
The generator is trained to produce samples that are close to the true posterior distribution, while the discriminator is trained to distinguish between true and generated samples. 
These two networks are trained in an adversarial fashion, where the generator tries to fool the discriminator and the discriminator tries to distinguish between true and generated samples. 
This adversarial process pushes both networks to improve, leading to generators that can provide high quality samples. 

Generative conditional modelling, such as cGANs, offer a powerful framework for radio interferometric imaging, enabling uncertainty quantification and generation of diverse image samples that are consistent with observed data.
The goal is to train a generating function $\boldsymbol{G}_\theta: \boldsymbol{Z} \times \boldsymbol{Y} \rightarrow \boldsymbol{X}$ that maps from the observed data $\boldsymbol{y} \in \boldsymbol{Y} \in \mathbb{C}^M$ and a latent code vector $\boldsymbol{z} \in \boldsymbol{Z} \sim \mathcal{N}(\boldsymbol{0}, \boldsymbol{I})$ to an image $\hat{\boldsymbol{x}} \in \boldsymbol{X} \in \mathbb{R}^N$ such that $\hat{\boldsymbol{x}} = \boldsymbol{G}_\theta(\boldsymbol{z}, \boldsymbol{y}) \sim p_{{\hat{\boldsymbol{x}}} \mid \boldsymbol{y}}(\cdot \mid \boldsymbol{y})$, with the generated posterior distribution $p_{{\hat{\boldsymbol{x}}} \mid \boldsymbol{y}}(\cdot \mid \boldsymbol{y})$ being as close as possible to the true posterior distribution of the image given the observed data $p_{{\boldsymbol{x}} \mid \boldsymbol{y}}(\cdot \mid \boldsymbol{y})$. 
Additionally, one also trains a discriminator function $\boldsymbol{D}_\phi: \boldsymbol{X} \rightarrow [0,1]$ that tries to distinguish between true samples $\boldsymbol{x} \sim p_{{\boldsymbol{x}} \mid \boldsymbol{y}}(\cdot \mid \boldsymbol{y})$ and generated samples $\hat{\boldsymbol{x}} \sim p_{{\hat{\boldsymbol{x}}} \mid \boldsymbol{y}}(\cdot \mid \boldsymbol{y})$.

However, GANs, and subsequently cGANs, suffer from training instability and mode collapse \citep{arjovskyPrincipledMethodsTraining2017}, where the generator produces samples with low diversity. 
Using the original GAN loss function \cite{goodfellowGenerativeAdversarialNetworks2014}, when the discriminator gets too good at distinguishing between the distributions of true and generated samples, the gradient of the generator vanishes and the training becomes unstable. 
In order to combat this, Wasserstein GANs (WGANs) have been introduced \citep{arjovskyWassersteinGenerativeAdversarial2017}, which use the Wasserstein-1 distance to measure the distance between the true and generated samples. 
The dual formulation of the Wasserstein-1 distance is given by
\begin{equation}\label{eq:Wasserstein}
    W_1\left( p_{{{\boldsymbol{x}}} \mid \boldsymbol{y}}(\cdot \mid \boldsymbol{y}), p_{\hat{\boldsymbol{x}} \mid \boldsymbol{y}}(\cdot \mid \boldsymbol{y}) \right)=\sup _{D_\phi \in  \text{ Lip}_1} \mathbb{E}_{\boldsymbol{x} \mid \boldsymbol{y}}\{D_\phi(\boldsymbol{x}, \boldsymbol{y})\} - \mathbb{E}_{\widehat{\boldsymbol{x}} \mid \boldsymbol{y}}\{D_\phi(\hat{\boldsymbol{x}}, \boldsymbol{y})\},
\end{equation}
where we have a discriminator that is a 1-Lipschitz function ($D_\phi \in \text{Lip}_1$) and $\hat{\boldsymbol{x}} = G_\theta(\boldsymbol{z}, \boldsymbol{y})$ is the generated image. 
The loss function of the WGAN generator can be found by taking the expectation over several measurements $\boldsymbol{y} \in \boldsymbol{Y}$ and latent code vectors $\boldsymbol{z} \in \mathcal{N}(0,1)$ and is given by
\begin{equation}\label{eq:adv-loss}
    \mathcal{L}_{\text {adv}}(\theta, \phi) \triangleq \mathbb{E}_{\boldsymbol{x}, \boldsymbol{z}, \boldsymbol{y}}\left\{ D_{\phi}(\boldsymbol{x}, \boldsymbol{y}) - D_{\phi}\left(G_{\theta}(\boldsymbol{z}, \boldsymbol{y}), \boldsymbol{y}\right)\right\},
\end{equation}
which approximates the Wasserstein-1 distance. Through training the discriminator, the quality of the Wasserstein approximation improves.
The loss function for the discriminator is thus given by
\begin{equation}\label{eq:discrim-loss}
    \mathcal{L}_{\text {discrimator}}({\theta}, {\phi}) = - \mathcal{L}_{\text {adv}}({\theta}, {\phi}) +  \mathcal{L}_{\text {gp}}(\phi),
\end{equation}
where $\mathcal{L}_{\text {gp}}(\phi)$ is a gradient penalty term that ensures the discriminator remains 1-Lipschitz \citep{gulrajaniImprovedTrainingWasserstein2017a} and the loss is minimised with respect to the discriminator parameters $\phi$. 
The expectation value over $\boldsymbol{x},\boldsymbol{y}$ can be expressed as an average over the training pairs $\left\{  \boldsymbol{x}_i,\boldsymbol{y}_i \right\}$ and the weights of the generator and discriminator networks are optimised alternatingly by minimising the loss functions.

Besides the training instability and mode collapse found in unconditional GANs, conditional GANs can also suffer an additional cause of mode collapse. 
Since there is only one true image $\boldsymbol{x}_i$ for each observed measurement $\boldsymbol{y}_i$, there is no incentive for the generator to produce diverse samples. 
To prevent this, \citet{bendelRegularizedConditionalGAN2023} introduced a regularised cGAN (rcGAN) that uses several regularisation terms to ensure diversity in the generated samples. 
Instead of minimising Equation~\ref{eq:adv-loss}, the generator is trained by minimising
\begin{equation}\label{eq:rcGAN-loss}
    \underset{{\theta}}{\arg \min} \left\{\beta_{\text {adv }} \mathcal{L}_{\text {adv }}({\theta}, {\phi}) + \mathcal{L}_{1, N_{\text {train }}}({\theta})-\beta_{\mathrm{SD}} \mathcal{L}_{\mathrm{SD}, N_{\text {train }}}({\theta}) \right\},
\end{equation}
where $\beta_{\text {adv }} > 0$ and $\beta_{\mathrm{SD}} > 0$ are hyperparameters and $N_{\text {train }}>2$ is the number of samples used in the regularisation terms. 
The $\mathcal{L}_{1, N_{\text {train }}}(\boldsymbol{\theta})$ regularisation term is an $N_{\text{train}}$ supervised $\ell_1$ loss
\begin{equation}
    \mathcal{L}_{1, N_{\text {train }}}(\boldsymbol{\theta}) \triangleq \mathrm{E}_{\mathrm{x}, \mathbf{z}_1, \ldots, \mathbf{z}_n, \mathrm{y}}\left\{\left\|\boldsymbol{x}-\widehat{\boldsymbol{x}}_{\left(N_{\text {train }}\right)}\right\|_1\right\}, \\
\end{equation}
with $\widehat{\boldsymbol{x}}_{\left(N_{\text{train}}\right)} = \sum_{i=1}^{N_{\text {train }}} \frac{\boldsymbol{G_\theta}(\boldsymbol{z}_i, \boldsymbol{y})}{N_{\text{train}}} $ the $N_{\text{train}}$-sample average and $\widehat{\boldsymbol{x}}_i \triangleq \boldsymbol{G_\theta}(\boldsymbol{z}_i, \boldsymbol{y})$ the generated samples.

The $\mathcal{L}_{\mathrm{SD}, N_{\text {train }}}(\boldsymbol{\theta})$ term is a standard deviation reward that encourages the samples to deviate by looking at the $\ell_1$ distance between the individual samples and the $N_{\text{train}}$-sample average
\begin{equation}
    \begin{aligned}
        \mathcal{L}_{\mathrm{SD}, N_{\text {train }}}(\boldsymbol{\theta}) & \triangleq \sqrt{\frac{\pi}{2 N_{\text {train }}\left(N_{\text {train }}-1\right)}} \\
        & \sum_{i=1}^{N_{\text {train }}} \mathrm{E}_{\mathbf{z}_1, \ldots, \mathrm{z}_p, \mathrm{y}}\left\{\left\|\widehat{\boldsymbol{x}}_i-\widehat{\boldsymbol{x}}_{\left(N_{\text {train }}\right)}\right\|_1\right\}.
    \end{aligned}
\end{equation}

In the work of \citet{bendelRegularizedConditionalGAN2023}, the authors show that this loss function results in generated samples that follow the true posterior distribution in both mean and covariance if the elements of $\hat{\boldsymbol{x}}_i$ and $\boldsymbol{x}$ are independent Gaussian distributions conditioned on $\boldsymbol{y}$ \citep[][\S Prop. 3.1]{bendelRegularizedConditionalGAN2023}:
\begin{align}
    \label{eq:point-estimate}
    \mathbb{E}_{\boldsymbol{z_i} | \boldsymbol{y}} \{ \hat{\boldsymbol{x}}_i(\theta^*) | \boldsymbol{y}  \} &= \mathbb{E}_{\boldsymbol{x} | \boldsymbol{y}} \left\{ \boldsymbol{x} | \boldsymbol{y} \right\} = \hat{\boldsymbol{x}}_{\text{mmse}} \\
    \label{eq:uncertainty-estimate}
    \text{Cov}_{\boldsymbol{z_i} | \boldsymbol{y}} \left\{ \hat{\boldsymbol{x}}_i(\theta^*) | \boldsymbol{y}  \right\} &= \text{Cov}_{\boldsymbol{x} | \boldsymbol{y}} \left\{ \boldsymbol{x} | \boldsymbol{y}  \right\}
\end{align}
In order to ensure that this standard deviation reward term results in generated samples following the true posterior distribution in both mean and covariance, the hyperparameter $\beta_{\text {SD}}$ needs to be tuned carefully, as choosing a larger value of $\beta_{\text {SD}}$ tends to result in samples with larger variance.

To automatically tune the $\beta_{\text {SD}}$ reward weight, \citet{bendelRegularizedConditionalGAN2023} propose a method that relates the error of a single generated sample to that of the error on the $N$-average of the samples. 
The authors show that for independent samples of the true posterior these errors should follow the relation
\begin{equation}
    \frac{\mathcal{E}_1}{\mathcal{E}_{\text{N}}} = \frac{2 N}{N+1},
\end{equation}
where the errors are calculated over the samples of the validation set as 
\begin{equation}
    \begin{aligned}
        \widehat{\mathcal{E}_N} = \frac{1}{N_{\text{val}}} \sum_{i=1}^{N_{\text{val}}} \left\| \boldsymbol{x}_i - \frac{1}{N_{\text{val}}} \sum_{j=1}^{N_\text{val}}\boldsymbol{G}_\theta(\boldsymbol{z}_j, \boldsymbol{y}_i) \right\|_1, \\ 
        \widehat{\mathcal{E}_1} = \frac{1}{N_{\text{val}}} \sum_{i=1}^{N_{\text{val}}} \left\| \boldsymbol{x}_i - \boldsymbol{G}_\theta(\boldsymbol{z}_1, \boldsymbol{y}_i) \right\|_1. 
    \end{aligned}
\end{equation}
The authors then use a gradient descent method to update the $\beta_{\text {SD}}$ hyperparameter using
\begin{equation}
    \beta_{\text {SD}, t+1} = \beta_{\text{SD}, t} - \mu_{\text{SD}} \cdot \left( \log_{10} \left[ \frac{\widehat{\mathcal{E}_1}}{\widehat{\mathcal{E}_{\text{N}}}}\right] - \log_{10}\left[\frac{2 N}{N+1}\right]\right) \cdot \beta_{\text{SD}, 0},
\end{equation}
with some $\mu_{SD} > 0$ as the learning rate. This update is performed after every epoch using the validation set to ensure that the generated samples follow the true posterior in covariance. 
For a more in-depth discussion on the tuning of the hyperparameters and proof of this proposition we refer the reader to \citet{bendelRegularizedConditionalGAN2023}.

\section{Generative radio interferometric imaging}\label{sec:proposed-methods}
\begin{figure*}
    \centering
    \includegraphics[width=\textwidth, trim= 0 0cm 0 0, clip]{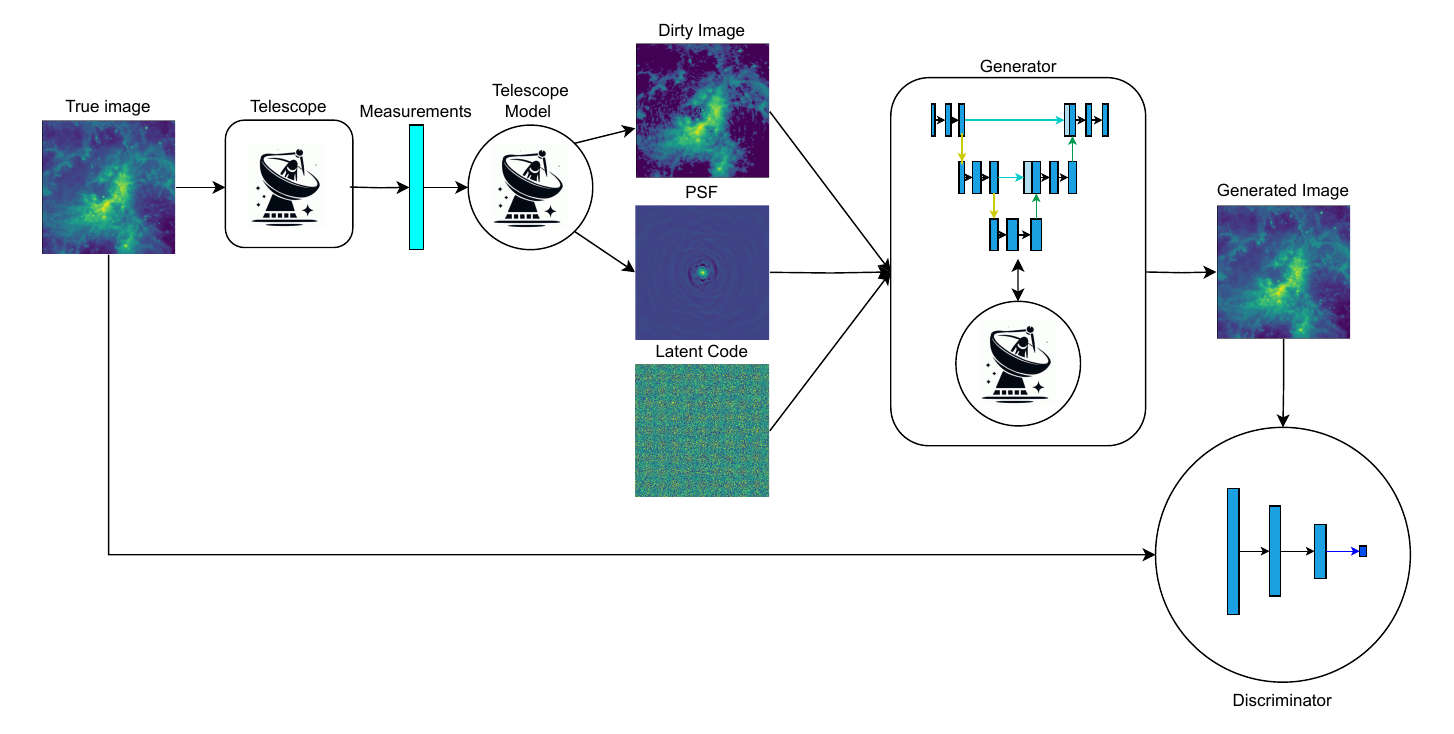}
    \caption[A diagram of the rcGAN framework.]{
        Our RI-GAN framework consisting of the generator and discriminator networks. From the true image we simulate measurements from which we can create the dirty image and PSF for an observation. These two images, combined with a latent code image, are used as input to the generator network. The generator network outputs an approximate sample from the posterior distribution. The discriminator gets passed either a generated image or a true image and outputs a scalar.}
    \label{fig:framework}
\end{figure*}

In this work we build on the rcGAN framework of \citet{bendelRegularizedConditionalGAN2023}, adapt it to the radio interferometry problem and extend it by explicitly including the measurement operator in the generator network. A schematic of our RI-GAN framework is shown in Figure~\ref{fig:framework}.

In order to provide a model that addresses the scalability problem for observations with large telescopes like the SKA, the models work using just the dirty image and PSF of the observation in order to reduce the number of evaluations of the measurement operator. These need to only be computed once per observation, which is significantly less than the number of evaluations needed for iterative optimisation approaches.

\subsection{Generator}
The generator network is based on the U-Net architecture \citep{ronnebergerUNetConvolutionalNetworks2015} which has been shown to be effective in learned post-processing settings in radio interferometric imaging \citep[e.g.][]{allamjrRadioInterferometricImage2016,terrisDeepPostProcessingSparse2019, marsLearnedInterferometricImaging2023b,marsLearnedRadioInterferometric2025} and has been used extensively for GANs. 
The input of the generator consists of several image channels:
\begin{enumerate}
    \item The dirty image, $\boldsymbol{x}_{\text{dirty}} = \boldsymbol{\Phi}^\dagger \boldsymbol{y}$, allowing the generator to condition on the observed data. 
    \item The PSF of the telescope, which is used to condition the generator on the measurement operator that has a different PSF for each \emph{uv}-coverage. 
    \item A latent code image of random noise. By varying this latent code image, we generate different posterior samples. 
\end{enumerate}
The output of the generator is a sample from the approximate posterior distribution.
The U-Net architecture consists of four layers with strided convolutions to downsample from each encoder layer to the next and transposed convolutions to upsample from each decoder layer to the next. Each encoder and decoder layer consists of residual convolutional blocks, which are based on the implementation in \citet{AdlerÖktem+2025+359+412}, with skip connections from the encoder to the decoder layers. A more detailed description of the architecture can be found in the paper of the rcGAN framework \citep{bendelRegularizedConditionalGAN2023}.

To enhance the robustness and quality of the reconstruction, we incorporate the explicit use of the measurement operator into the generator network. For this, we consider a GU-Net \citep[][]{marsLearnedInterferometricImaging2023b,marsLearnedRadioInterferometric2025} architecture, which integrates the gradient of the data-fidelity likelihood, $\mathcal{L} = || \boldsymbol{y} - \boldsymbol{\Phi} \boldsymbol{x} ||_{\ell_2}^2$, into the network. At several layers of the U-Net architecture, we compute the gradient of the data-fidelity likelihood with respect to the image $\boldsymbol{x}$ and combine it with the rest of the channels using convolutional layers. 
This gradient of the data-fidelity likelihood is given by 
\begin{equation}
    \nabla_{\boldsymbol{x}} \mathcal{L} \propto \boldsymbol{\Phi}^* (\boldsymbol{y} - \boldsymbol{\Phi} \boldsymbol{x}).
\end{equation}

Given the computational cost of the full measurement operator with (de)gridding operations (as well as w-stacking and w-correction operations in case of non-coplanar baselines), we approximate the $\boldsymbol{\Phi}^* \boldsymbol{\Phi} $ operation by a convolution with the PSF of the telescope.
This approximation is accurate in the case of coplanar baselines, though it neglects wide-field effects caused by non-coplanar baselines. 
This approximate gradient is computed at full resolution scale at the first level of the U-Net and lower resolution gradients are applied on the lower levels of the network:
\begin{equation}
    \widetilde{\nabla}_{x,i} \mathcal{L} \propto \boldsymbol{x}_{\text{dirty}, i} - \boldsymbol{x}_i * \boldsymbol{I}_{\text{PSF,} i},
\end{equation}
where the subscript $i$ denotes the $i$-th level in the network. The dirty image $\boldsymbol{x}_{\text{dirty}, i}$ and the PSF $\boldsymbol{I}_{\text{PSF,} i}$ are sub-sampled versions of the full resolution dirty image and PSF, sub-sampled by a factor of $2^i$ along each dimension to match the resolution of the channels on the $i$-th level. 
These (sub-scale) gradient operations are applied at the beginning of each encoder and decoder layer in the network.

\subsection{Discriminator}
The discriminator network is a standard convolutional neural network (CNN) consisting of one convolutional layer followed by six convolutional downsampling blocks. 
Each block consists of a $2\times2$ average pool layer and a convolutional layer with instance normalisation and leaky ReLU activation functions (slope 0.2). 
The input of the discriminator is either an image from the true distribution or a generated image from the generator network. 
The output of the discriminator is a single scalar that represents the estimated Wasserstein score for the image. 

\subsection{Training data and training}
To train the generative networks, we compiled a large dataset of 13,000 simulated radio interferometric measurements of $360 \times 360$ galaxy images created from the IllustrisTNG simulations \citep{nelsonIllustrisTNGSimulationsPublic2019}. 
The images are created by binning and in-painting the image from the simulated particles, as described in more detail in \citet{marsLearnedInterferometricImaging2023b}. 
The measurements are simulated for different \emph{uv}-coverages of the MeerKAT telescope with a total of 241,920 visibilities, where we vary the pointing of the telescope. However, we assume co-planar baselines and neglect the $w$-components of the baselines. 
The added noise is calculated according to the Brown Equation \citep{brownTechnicalSpecificationMillimeter1998a} as implemented in the casatools simulation module that is part of CASA \citep{thecasateamCASACommonAstronomy2022}.
From the contaminated measurements, we create the dirty images and PSFs using Briggs weighting (using a robustness parameter of 0.5). 
For training, we use 10,000 images, with 1,500 images for validation and testing each.

The RI-GANs are trained for 100 epochs on 4 NVIDIA A100 GPUs with 80Gb VRAM each, using an effective batch size of 16; training takes about 48 hours. 

\subsection{Point estimate and uncertainty estimation}
The trained generator network can be used to generate approximate posterior samples when given the dirty image and PSF of an observation. From these samples, we compute a point estimate by taking the mean over $N$ samples, where $N$ is the number of samples generated. In order to estimate the uncertainty of the point estimate, we compute the standard deviation of the samples. These choices are motivated by proposition 3.1 in \citet{bendelRegularizedConditionalGAN2023}, summarised in Equations~\ref{eq:point-estimate} and \ref{eq:uncertainty-estimate}. In order to reduce the training and validation time we use $N_{\text{train}} = 2$ and $N_{\text{val}} = 8$ samples for training and validation respectively and found no significant drop in training performance compared to training using a larger number of samples. During evaluations we want to generate a larger number of samples in order to get a better estimate of the reconstruction and its uncertainty, so we use $N_{\text{eval}} = 32$ samples for the evaluation of the model. Note that for creating a reconstruction of a single observation, these samples can be generated in parallel on one or a few GPUs, which makes the generation of samples for a single observation very fast.

\section{Image reconstruction with RI-GAN}\label{sec:experiments}
The performance of the U-Net and GU-Net RI-GAN models is assessed based on reconstruction quality, uncertainty estimation, and computational efficiency. 
We evaluate these models using the same simulated data as described in the previous section and further validate them using simulated radio interferometric measurements from the 30 Doradus region. 
On the 30 Doradus region we also compare our reconstructions to the CLEAN algorithm \citep{offringaWSCleanImplementationFast2014}.

The quality of the image reconstructions is quantified using the signal-to-noise ratio:
\begin{equation}
    \text{SNR}(\boldsymbol{x}_{\text{true}}, \boldsymbol{x}_{\text{pred}}) = 20 \cdot \log_{10} \left( \frac{\left\| \boldsymbol{x}_{\text{true}} \right\|_2}{\left\| \boldsymbol{x}_{\text{true}} - \boldsymbol{x}_{\text{pred}} \right\|_2} \right).
\end{equation}

We also quantify the correlation of the uncertainty estimate with the absolute error of the reconstruction using the Pearson correlation coefficient:
\begin{equation}
    \text{corr}(\boldsymbol{x}_{\text{uncertainty}}, |\boldsymbol{x}_{\text{true}} - \boldsymbol{x}_{\text{pred}}|) = \frac{\text{cov}(\boldsymbol{x}_{\text{uncertainty}}, |\boldsymbol{x}_{\text{true}} - \boldsymbol{x}_{\text{pred}}|)}{\sigma_{\boldsymbol{x}_{\text{uncertainty}}} \sigma_{|\boldsymbol{x}_{\text{true}} - \boldsymbol{x}_{\text{pred}}|}},
\end{equation}
where $\text{cov}$ is the covariance, $\sigma$ is the standard deviation and $\boldsymbol{x}_{\text{uncertainty}}$ is the uncertainty estimate of the reconstruction.

\begin{figure*}
    \centering
    \includegraphics[scale=.44, trim=0.85cm 1.6cm 1cm 0.5cm, clip]{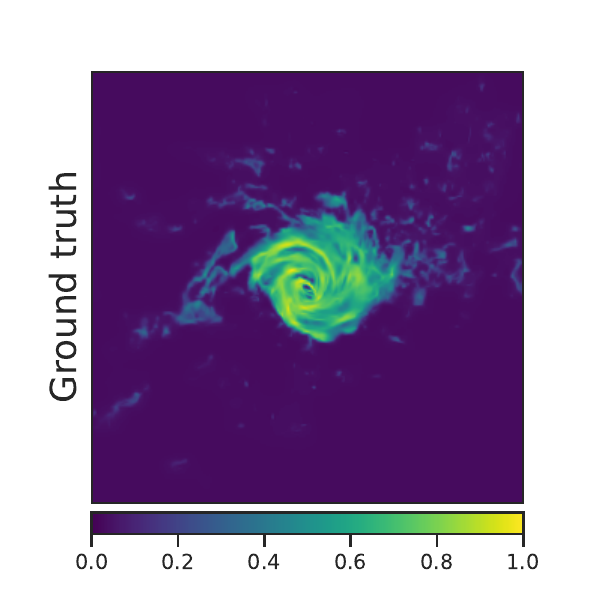}
    \includegraphics[scale=.44, trim=2.3cm 1.6cm 3cm 0.5cm, clip]{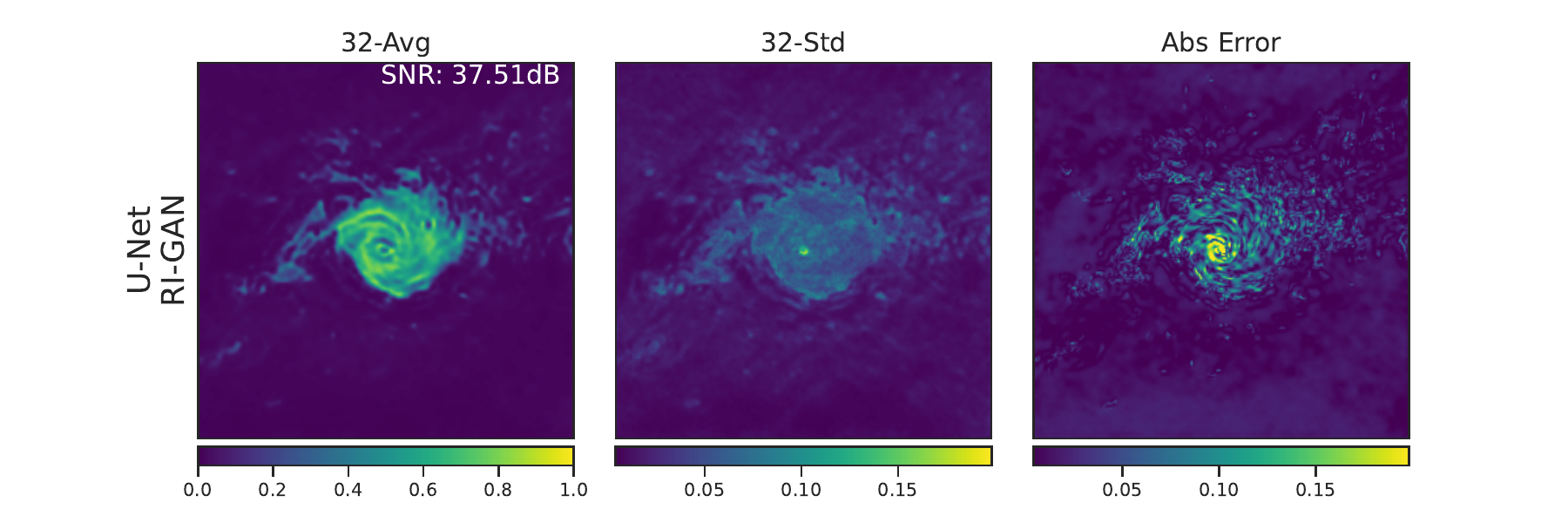}
    \includegraphics[scale=.44, trim=0.85cm .5cm 1cm 0.75cm, clip]{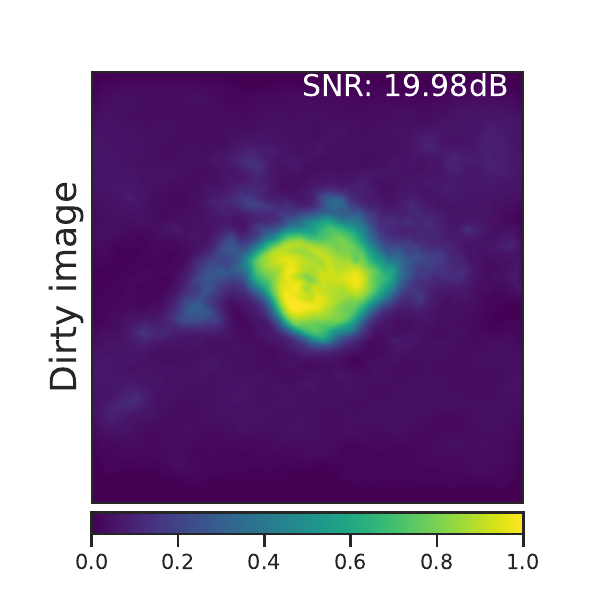}
    \includegraphics[scale=.44, trim=2.3cm .5cm 3cm 0.75cm, clip]{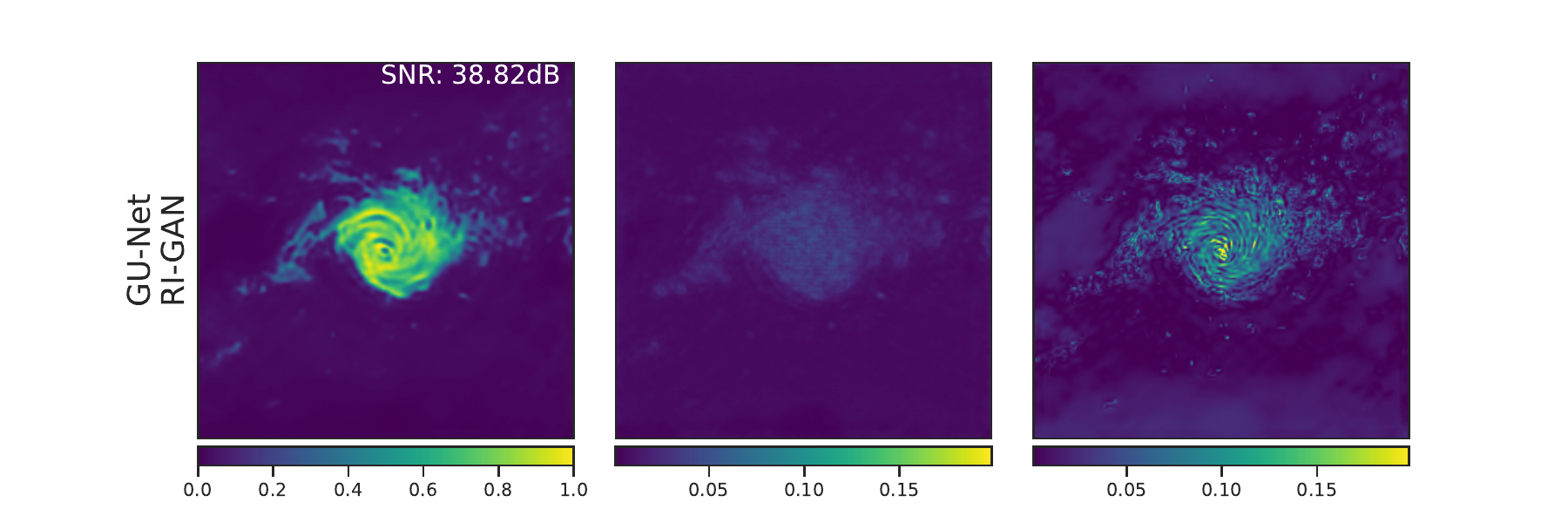}
    \includegraphics[scale=.44, trim=0.85cm 1.6cm 1cm 0.5cm, clip]{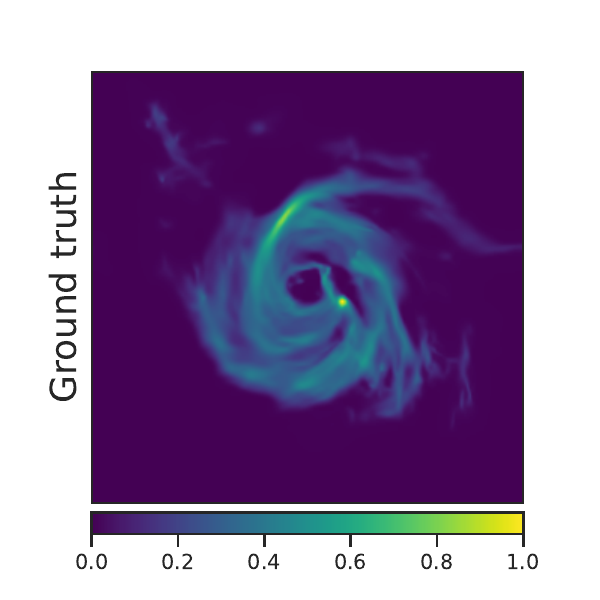}
    \includegraphics[scale=.44, trim=2.3cm 1.6cm 3cm 0.5cm, clip]{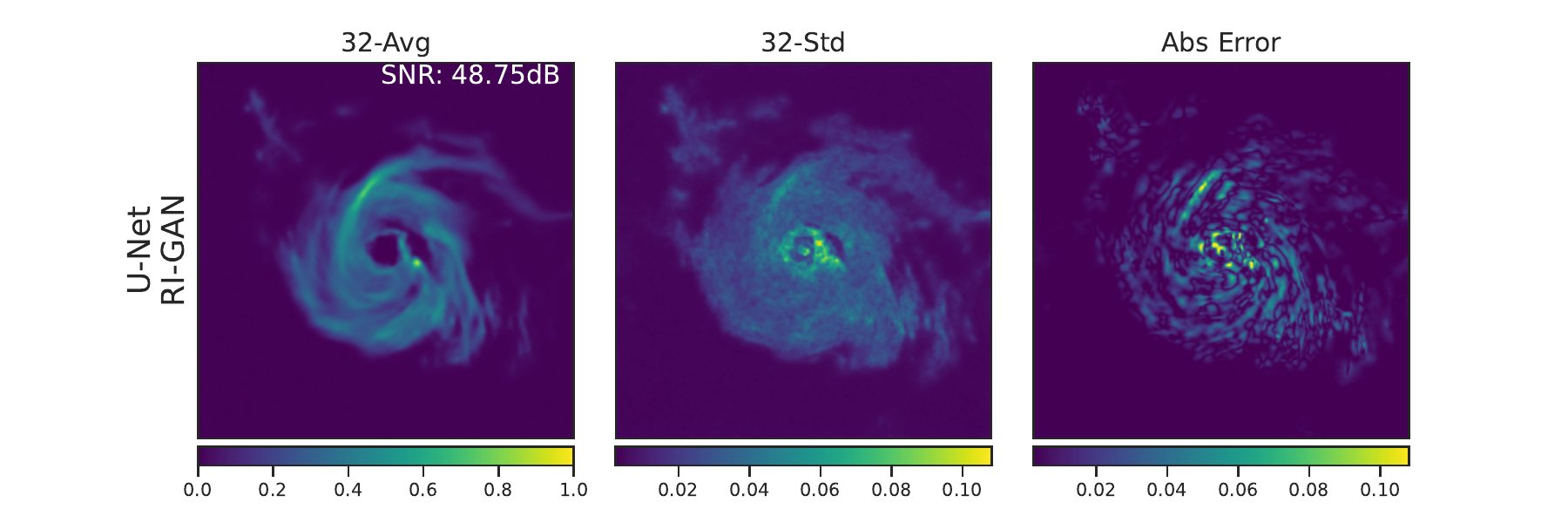}
    \includegraphics[scale=.44, trim=0.85cm .5cm 1cm 0.75cm, clip]{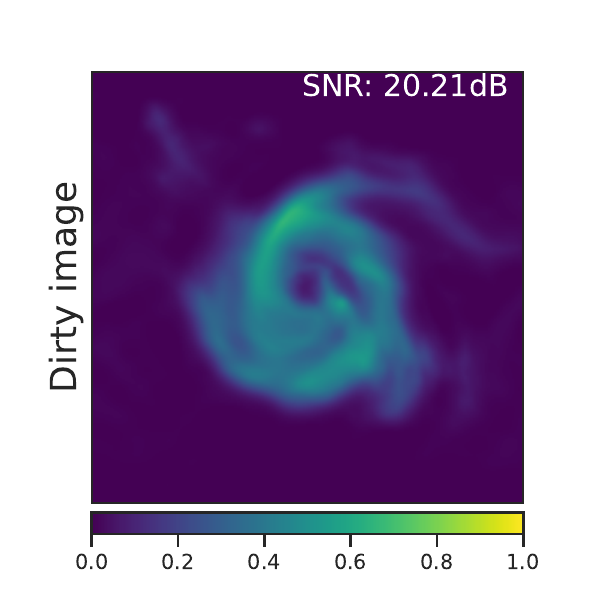}
    \includegraphics[scale=.44, trim=2.3cm .5cm 3cm 0.75cm, clip]{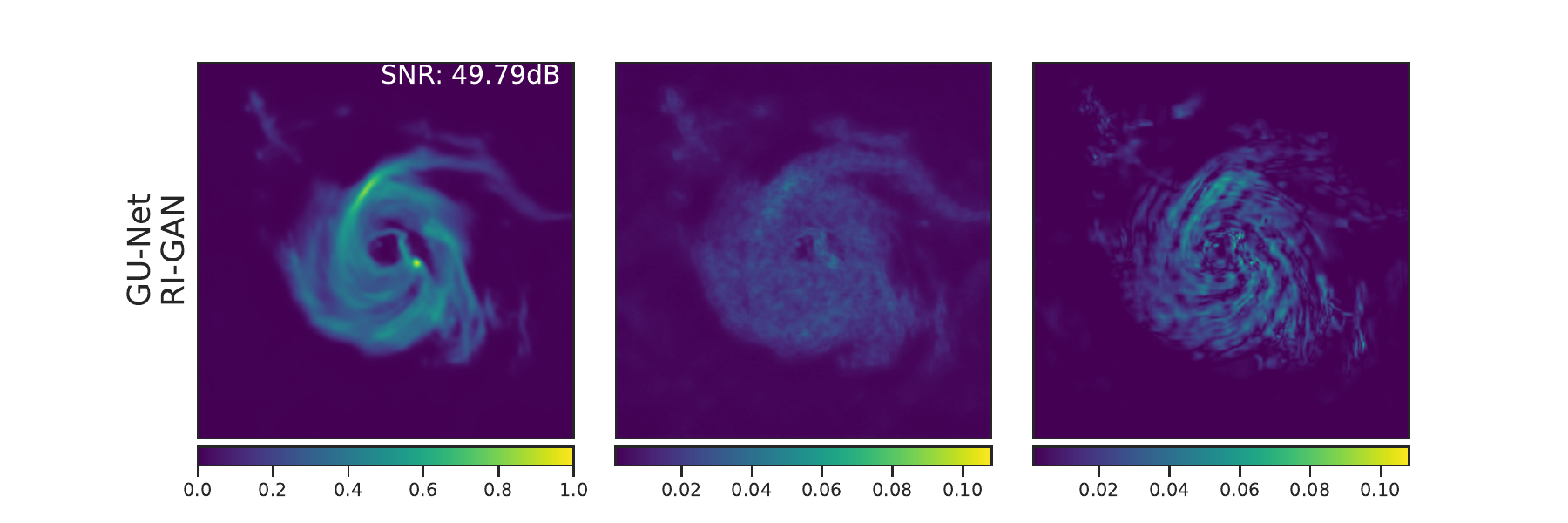}
    \includegraphics[scale=.44, trim=0.85cm 1.6cm 1cm 0.5cm, clip]{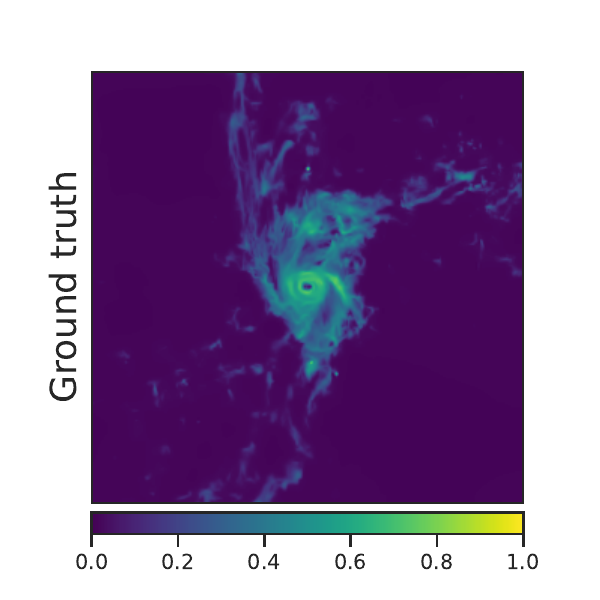}
    \includegraphics[scale=.44, trim=2.3cm 1.6cm 3cm 0.5cm, clip]{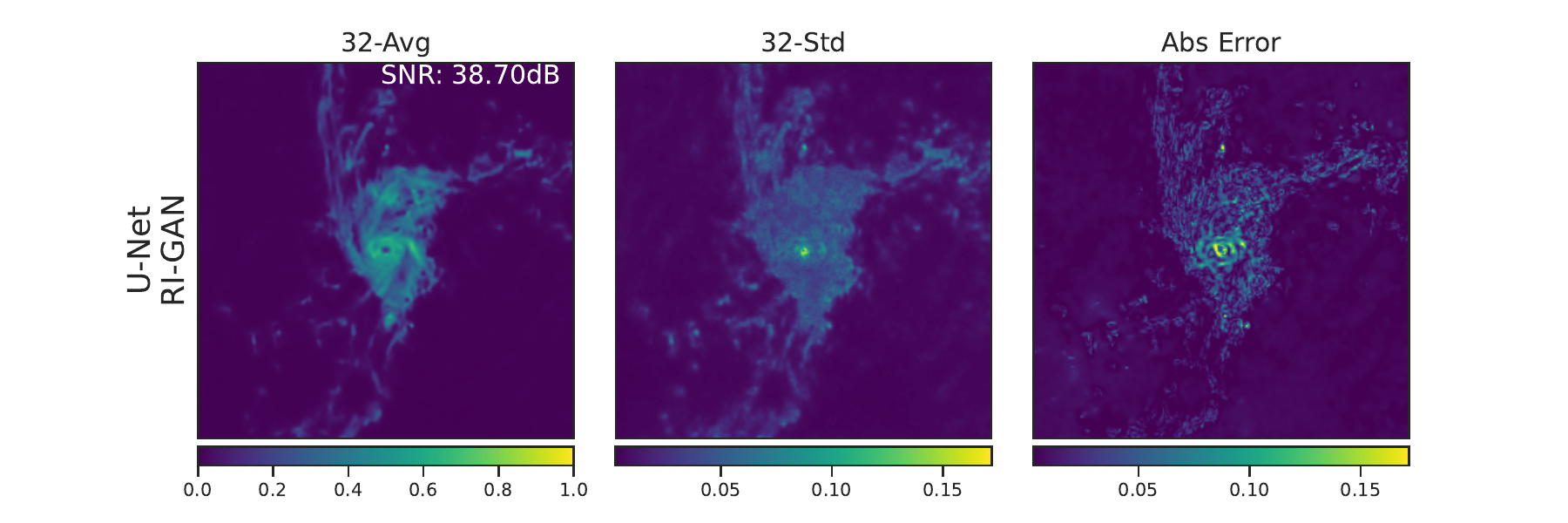}
    \includegraphics[scale=.44, trim=0.85cm .5cm 1cm 0.75cm, clip]{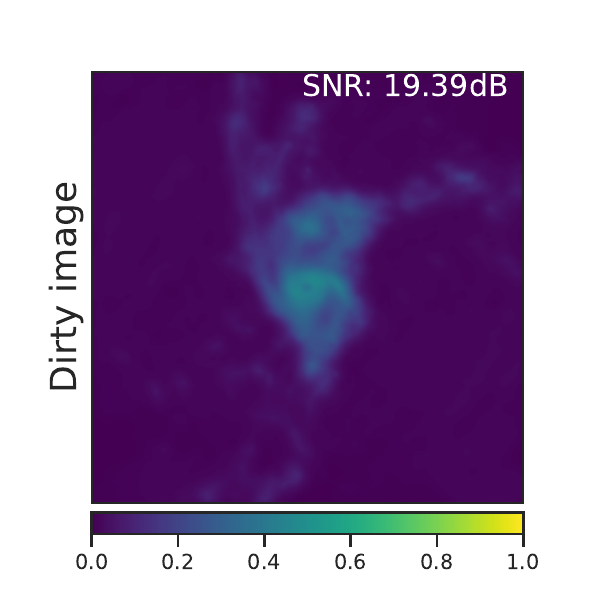}
    \includegraphics[scale=.44, trim=2.3cm .5cm 3cm 0.75cm, clip]{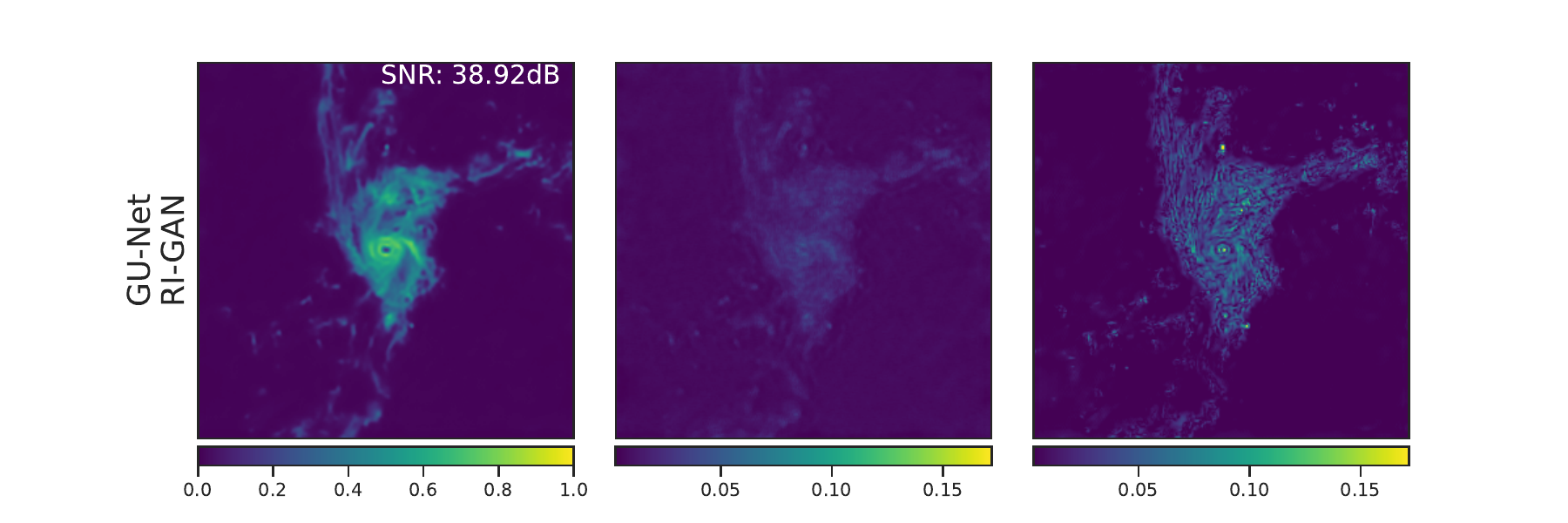}

    \caption[reconstructions using U-Net and GU-Net RI-GANs.]{
        Reconstructions using U-Net and GU-Net RI-GANs from simulated measurements of galaxies in test set. The reconstructions are calculated by averaging over 32 generated approximate posterior samples and the uncertainties are obtained by taking the standard deviation over these samples. Both models provide a large improvement over the dirty image. The GU-Net, which includes updates using an approximate gradient of the data fidelity term, provides higher quality reconstructions and lower uncertainties overall compared to the U-Net.}
    \label{fig:examples}
\end{figure*}

\subsection{Reconstruction quality}
\begin{figure}
    \centering
    \includegraphics[width=0.9\columnwidth, trim=0.8cm .5cm 2cm 1.5cm, clip]{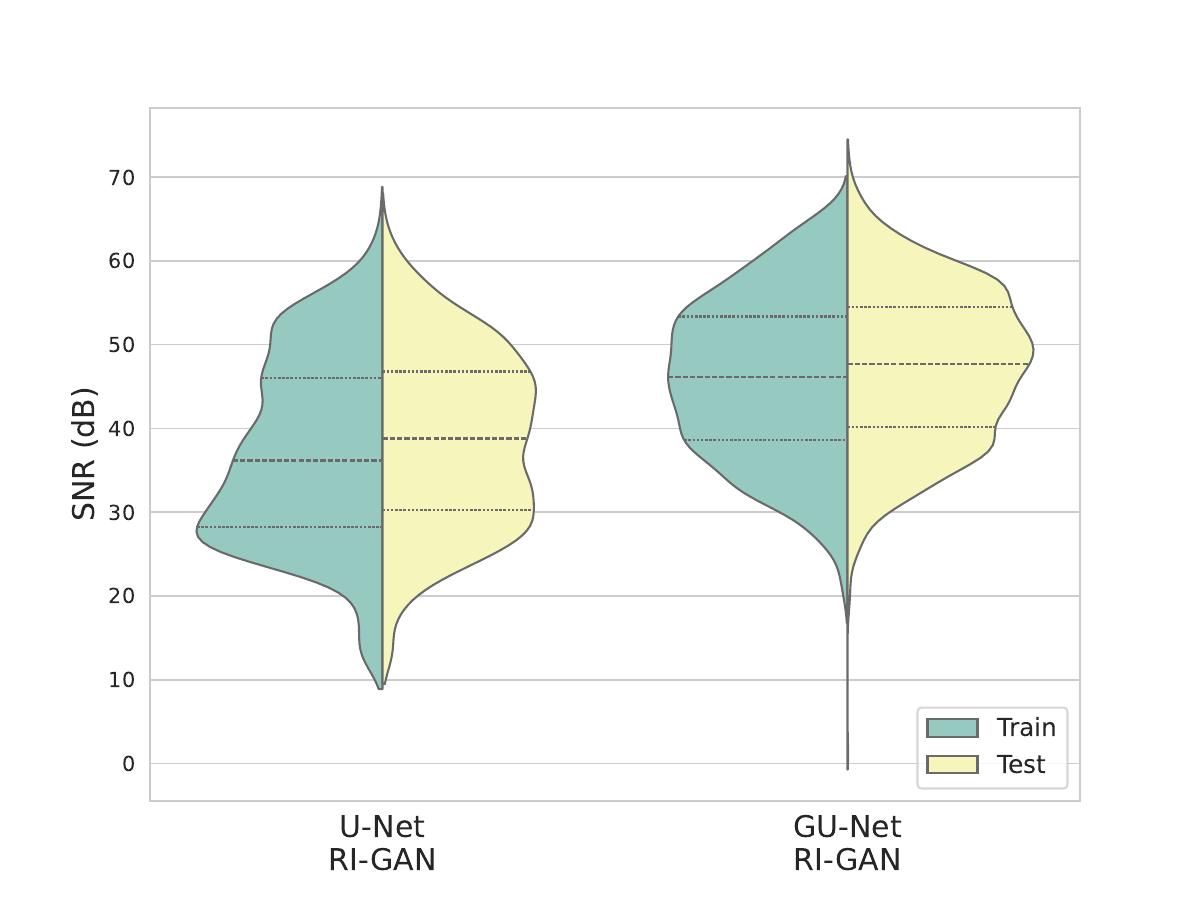}
    \caption[Distribution of the SNR for reconstructions using the U-Net and GU-Net RI-GANs.]{
        Distribution of the SNR for reconstructions using the U-Net and GU-Net RI-GANs for both the training and test set. The reconstructions are made by averaging over 32 generated posterior samples. Over both the train and test set, the GU-Net provides reconstructions with higher SNR compared to the U-Net. Besides that, the performance between the training and test set is similar for the GU-Net RI-GAN indicating that there is no overfitting to the train set.
        The dashed and dotted lines indicate the mean and quartiles of the distributions respectively.}
    \label{fig:violin}
\end{figure}

Figure~\ref{fig:examples} shows reconstructions using both the U-Net and GU-Net RI-GANs. 
The reconstructions are obtained by generating 32 posterior samples, with the mean representing the reconstruction and the standard deviation of these samples representing the reconstruction uncertainty. 
Both models reconstruct the images at high quality and the absolute error of the reconstructions visually correlates with the estimated uncertainty. 
The GU-Net consistently provides a higher average reconstruction quality and lower uncertainty compared to the U-Net. 
This trend is also evident in the SNR distribution across the dataset, as shown in Figure~\ref{fig:violin}.

\subsection{Number or samples}
\begin{figure}
    \centering
    \includegraphics[width=0.9\columnwidth, trim=3.2cm 0.2cm 20.5cm 1.5cm, clip]{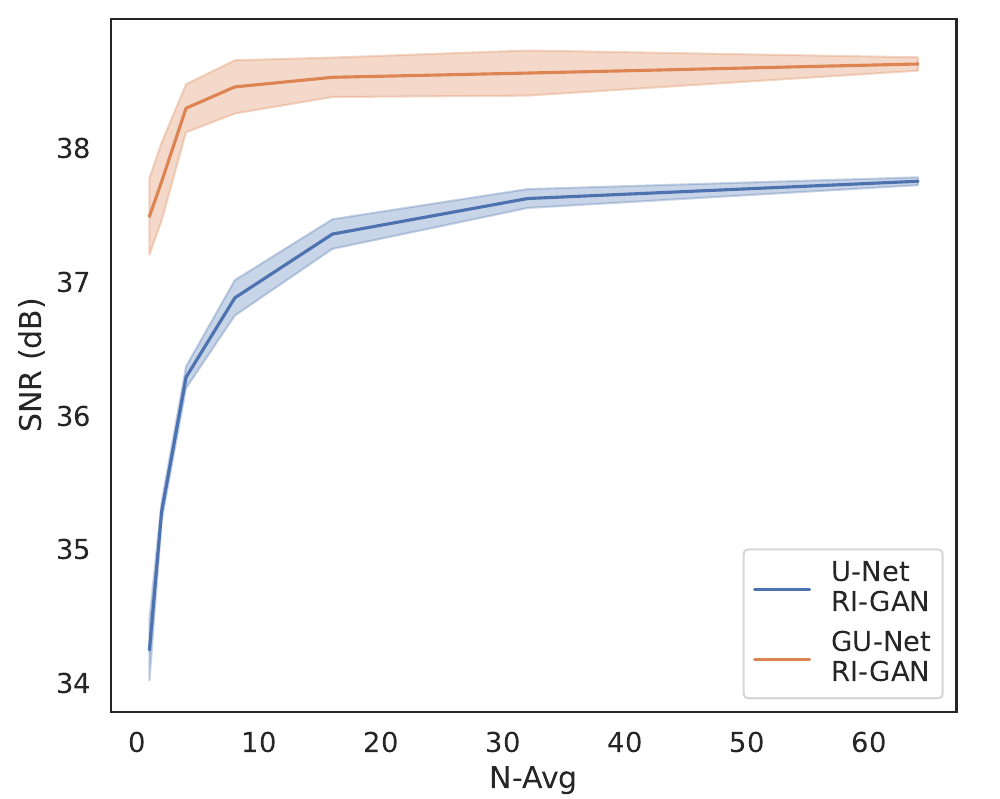}
    \includegraphics[width=0.9\columnwidth, trim=20.3cm 0.2cm 3.5cm 1.5cm, clip]{Figures/n_average_metrics_presentation.pdf}
    \caption[Effect of the number of samples on reconstruction quality and the correlation between the uncertainty and the absolute error for both the U-Net and the GU-Net RI-GANs.]{
        Effect of the number of samples on reconstruction quality and the correlation between the uncertainty and the absolute error for both the U-Net and the GU-Net RI-GANs for an example image. The shaded area represents the standard deviation of repeating the generation 100 times. Both the quality of the reconstruction as quantified by the SNR and the correlation of the uncertainty with the absolute error increase as a function of the number of samples. The GU-Net consistently outperforms the U-Net on both metrics.}
    \label{fig:n_average}
\end{figure}

The number of samples used to calculate the mean of the reconstructions affects the quality of the reconstructions and the uncertainties. 
During training and validation, we are able to use a smaller number of samples to speed up the training process, yet for the final reconstructions we want to use a larger number of samples to get the best estimate of the reconstruction and its uncertainty.
Figure~\ref{fig:n_average} shows the effect of increasing the number of samples for a particular reconstruction. 
As expected, increasing the number of samples used for the average improves the SNR of the reconstruction. 
This effect plateaus after 32 samples, indicating that the quality of the reconstruction does not significantly improve beyond this point. 
Additionally, the correlation between the estimated uncertainty also increases as a function of the number of samples, saturating around 32 samples at 0.58 for the U-Net and 0.69 for the GU-Net generators demonstrating that the uncertainty is a reasonable indicator of the error of the reconstruction.
The GU-Net reconstructions outperform the U-Net in both the SNR and the uncertainty correlation for any number of samples.

The observation that for a lower number of samples both the SNR and the uncertainty correlation are significantly lower than for higher number of samples indicates that there is significant diversity in the reconstructions, suggesting that we do not suffer from mode collapse. 

\subsection{Reconstruction of 30 Doradus}
\begin{figure*}
    \centering
    \includegraphics[scale=.7, trim=2.3cm 0cm 3cm -2cm, clip] {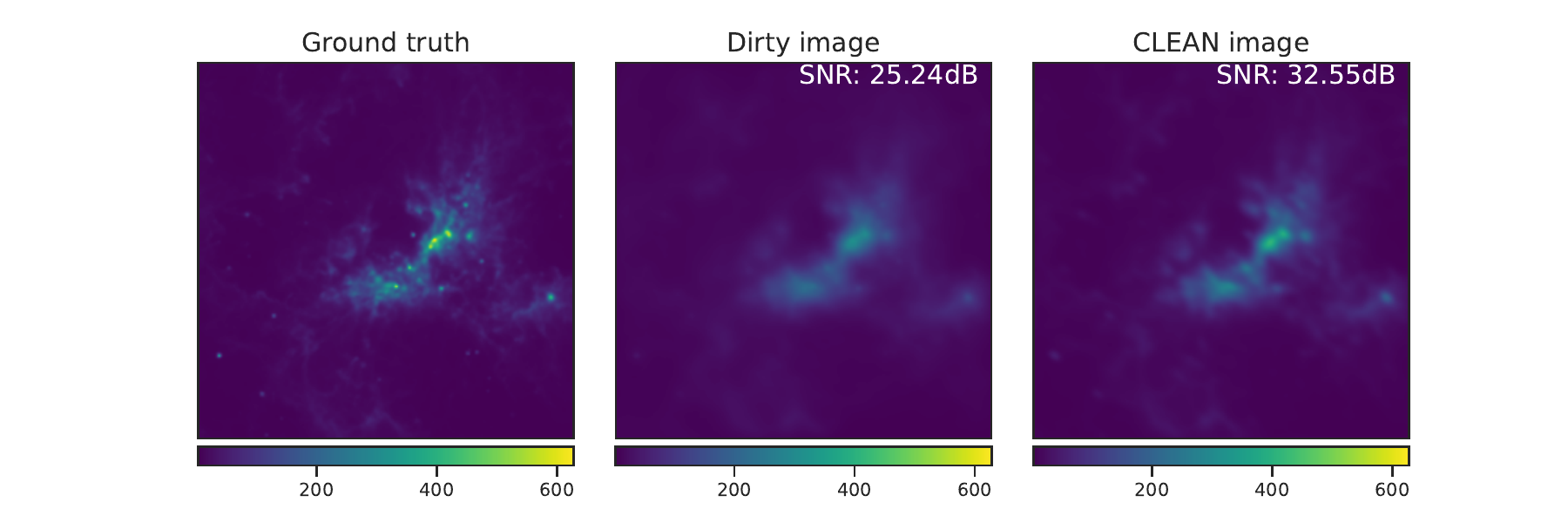}
    \includegraphics[scale=.7, trim=2.3cm 1.6cm 3cm 0.5cm, clip] {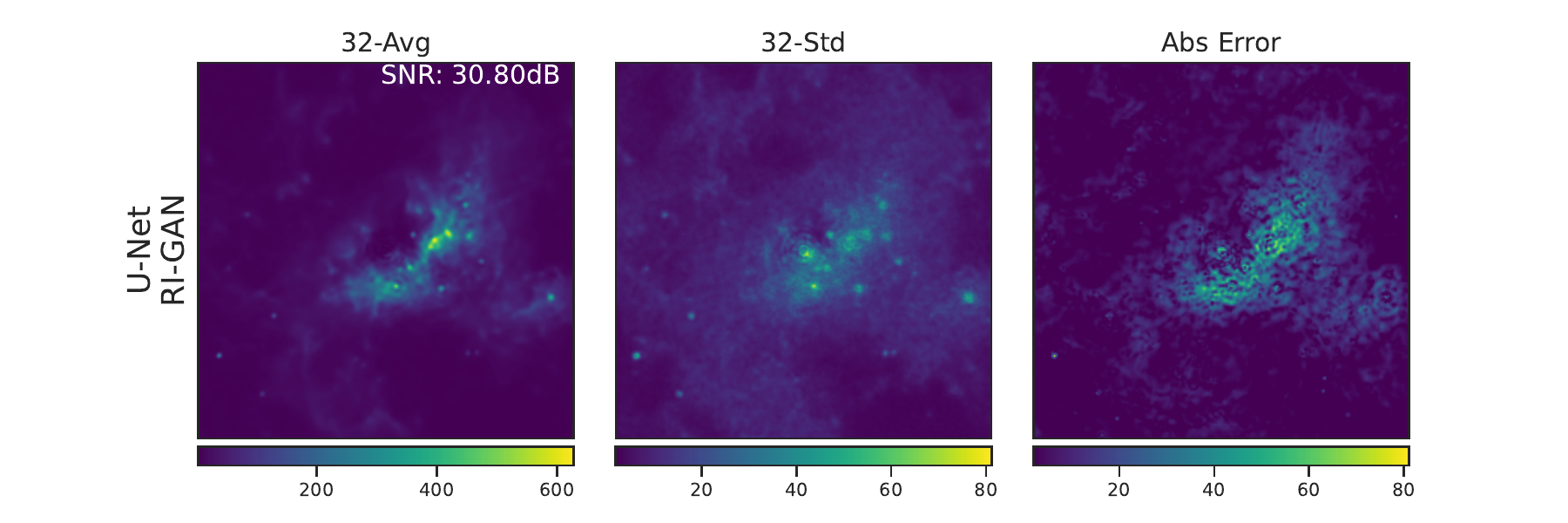}
    \includegraphics[scale=.7, trim=2.3cm .5cm 3cm 0.75cm, clip] {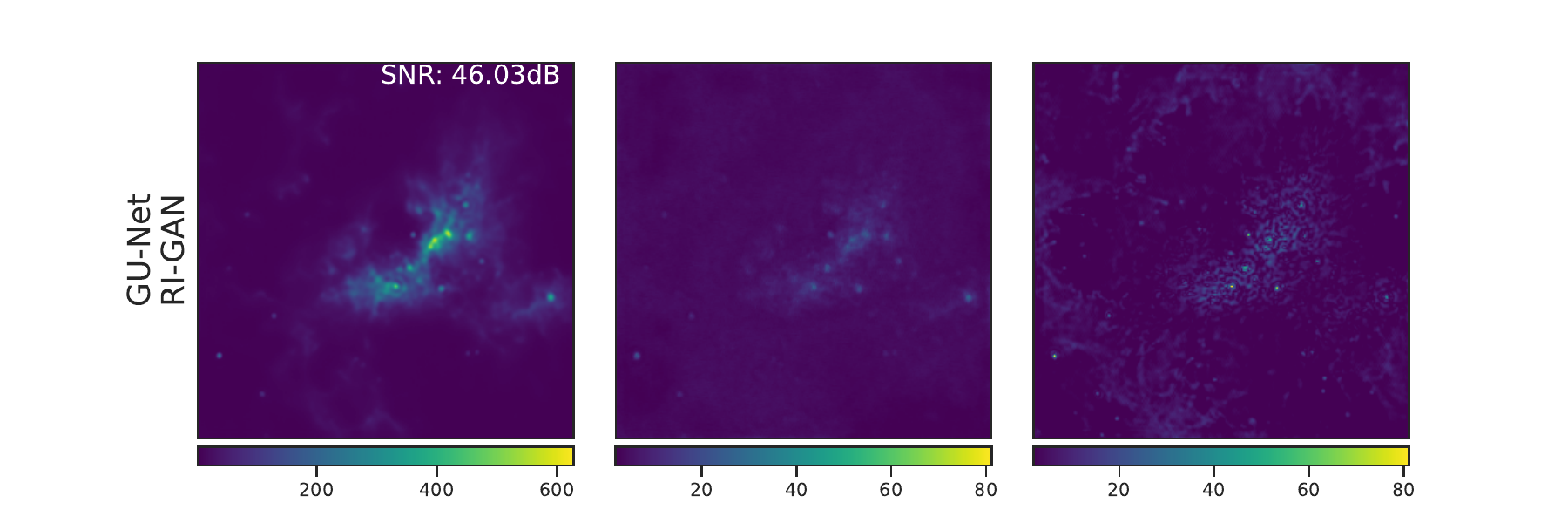}
    \caption[Reconstructions using U-Net and GU-Net RI-GANs from simulated measurements of the 30 Doradus region using an unseen \emph{uv}-coverage.]{
        Reconstructions using U-Net and GU-Net RI-GANs from simulated measurements of an out-of-distribution image of the 30 Doradus region using an unseen \emph{uv}-coverage. The reconstructions are calculated by averaging over 32 generated approximate posterior samples and the uncertainties are obtained by taking the standard deviation over these samples. The top row shows the ground truth, dirty image and the reconstruction using the CLEAN algorithm. The bottom rows show the reconstructions, uncertainties and absolute error using the U-Net and GU-Net RI-GANs respectively. All images are shown on a linear intensity scale. The GU-Net provides higher quality reconstructions and lower uncertainties compared to the U-Net and the CLEAN algorithm.}
    \label{fig:30Dor}
\end{figure*}

\begin{figure*}
    \centering
    \includegraphics[scale=.7, trim=2.3cm 0cm 3cm -2cm, clip] {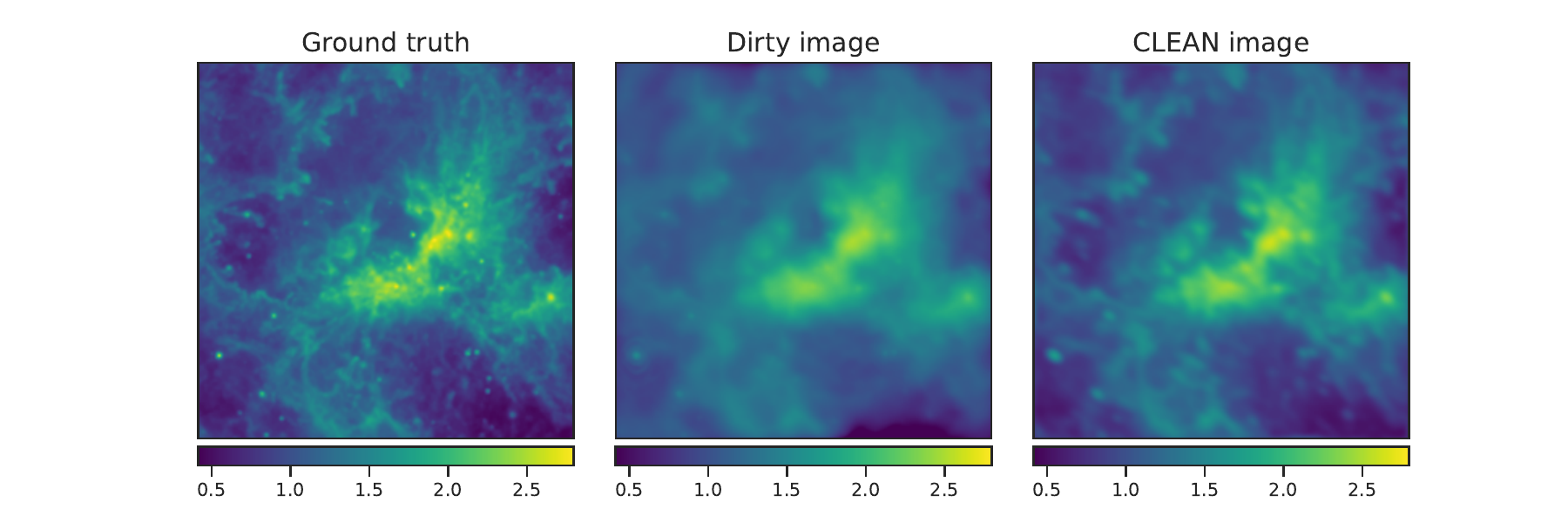}
    \includegraphics[scale=.7, trim=2.3cm 1.6cm 3cm 0.5cm, clip] {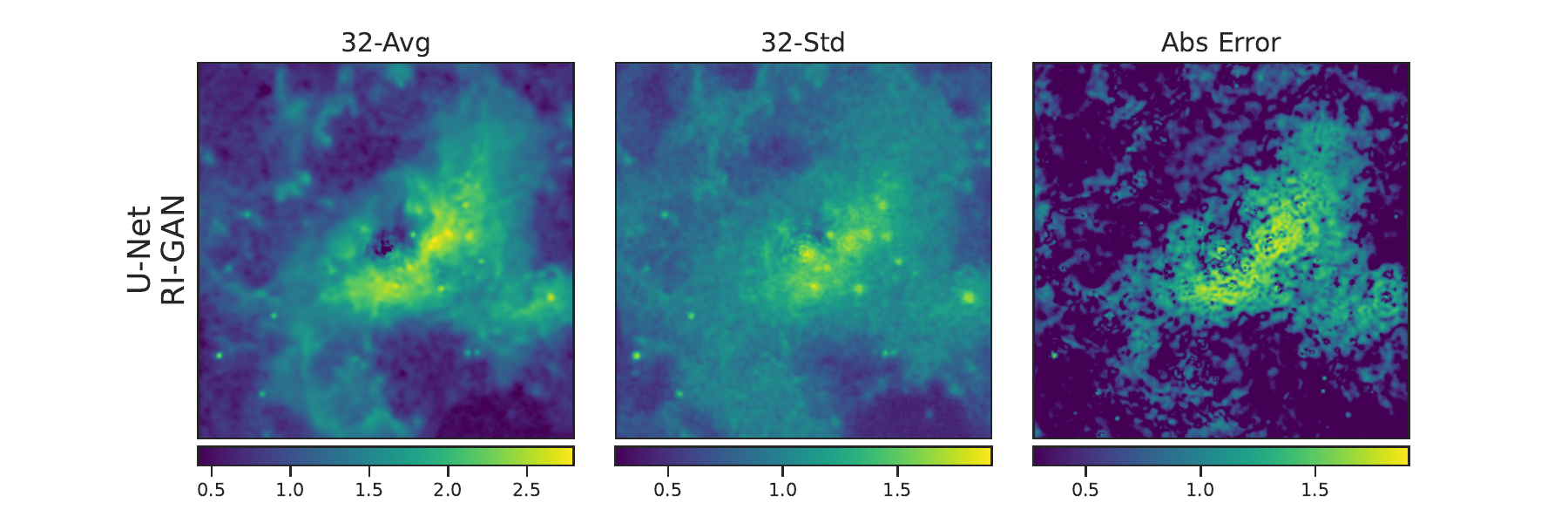}
    \includegraphics[scale=.7, trim=2.3cm .5cm 3cm 0.75cm, clip] {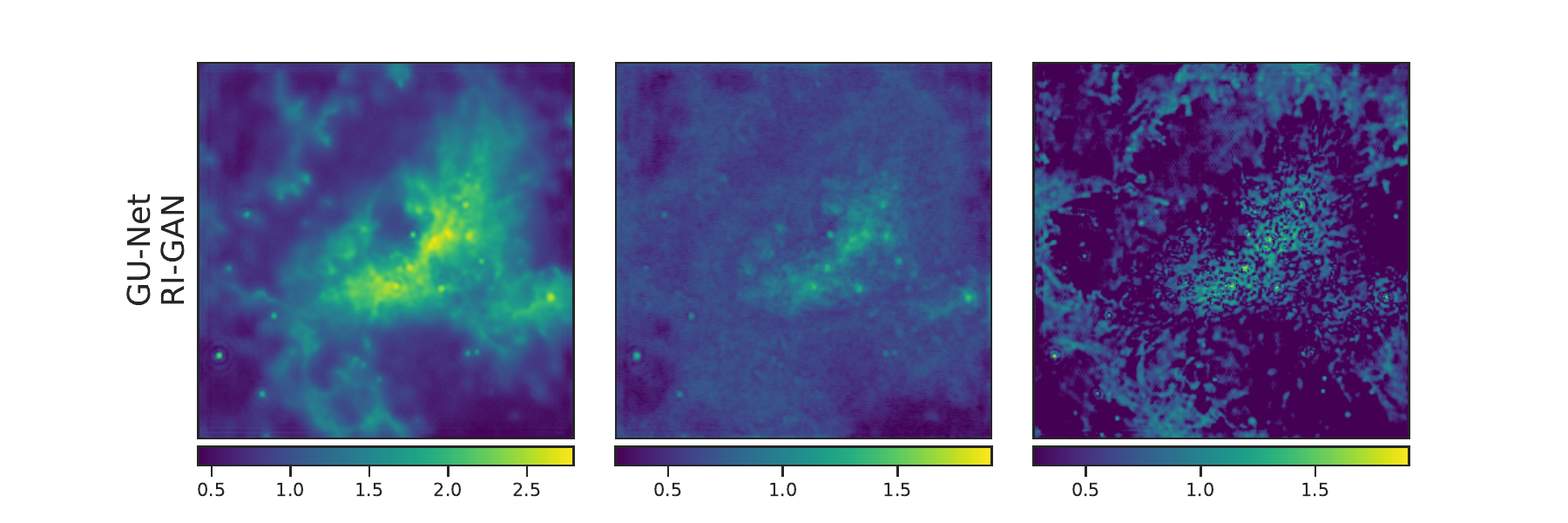}

    \caption[Reconstructions using U-Net and GU-Net RI-GANs from simulated measurements of the 30 Doradus region using an unseen \emph{uv}-coverage.]{
        Reconstructions using U-Net and GU-Net RI-GANs from simulated measurements of an out-of-distribution image of the 30 Doradus region using an unseen \emph{uv}-coverage. The reconstructions are calculated by averaging over 32 generated approximate posterior samples and the uncertainties are obtained by taking the standard deviation over these samples. The top row shows the ground truth, dirty image and the reconstruction using the CLEAN algorithm. The bottom rows show the reconstructions, uncertainties and absolute error using the U-Net and GU-Net RI-GANs respectively. All images are shown on a logarithmic intensity scale. The GU-Net provides higher quality reconstructions and lower uncertainties compared to the U-Net and higher resolution compared to the CLEAN algorithm.}
    \label{fig:30Dor_log}
\end{figure*}

To test the generators on out-of-distribution and more realistic data, we use an image of the 30 Doradus region, which has a larger dynamic range ($\sim 600$) than the simulated galaxies on which the models are trained (though not as high a dynamic range as real radio images which can be in the order of $10^6$). 
The measurements are simulated similarly to the training data but with a different, unseen \emph{uv}-coverage. 
Figures~\ref{fig:30Dor}~and~\ref{fig:30Dor_log} show the reconstructions on both a linear scale and a logarithmic scale to better showcase the performance at a higher dynamic range.
The GU-Net, which includes explicit evaluations of the measurement operator, improves reconstruction quality (SNR: $46.03$dB) compared to the U-Net (SNR: $30.80$dB), and the CLEAN algorithm (SNR: $32.55$dB).
The increased performance of the GU-Net is due to the explicit inclusion of the measurement operator in the generator, making it more robust to changes in the \emph{uv}-coverage and out-of-distribution data of a larger dynamic range. 
Also, note that the neural networks are trained on simulated data of a lower dynamic range and we expect that the performance of the GU-Net would improve further if it were trained on data more representative of the 30 Doradus region.

\pagebreak
\section{Conclusion}\label{sec:conclusion}
The use of generative models for interferometric imaging is a promising approach to provide fast image reconstructions of high quality as well as approximate uncertainty quantification, while being robust to different \emph{uv}-coverages. 
In this work we proposed our RI-GAN framework which builds on the rcGAN framework \citep{bendelRegularizedConditionalGAN2023}. 
These generative frameworks use a standard deviation regularisation term and automatic tuning of the regularisation strength to have the generated posterior samples follow the true posterior distribution in both mean and covariance. 
For our framework we explore both the use of a U-Net and a GU-Net generator \citep{marsLearnedInterferometricImaging2023b,marsLearnedRadioInterferometric2025}. 
The GU-Net generator is a modified U-Net generator, which includes explicit use of the measurement operator, by including at several points in the network an approximate gradient of an $L_2$-norm data fidelity term and using a convolution of the images with the point spread function (PSF). 
Adding the measurement operator explicitly in the reconstruction network has been shown to improve the quality of the reconstructions as well as provide robustness to varying visibility coverages \citep{marsLearnedRadioInterferometric2025}.

In our evaluation of the two models on measurements simulated from galaxies in the IllustrisTNG simulations, we find that the GU-Net RI-GAN consistently outperforms its U-Net counterpart in terms of image quality and shows a lower estimated uncertainty. 
When looking at the number of generated posterior samples needed for reconstruction and uncertainty estimation, we find that the quality of the reconstructions and the correlation of the uncertainties with the errors of the reconstructions increases with the number of samples, but saturates at approximately 32 samples. 
We also find that the inclusion of the measurement information explicitly in the generator increases the quality of the reconstruction and the uncertainties. 
For the test example we see that the correlation of the estimated uncertainty and the absolute error of the reconstruction is 0.69 for the GU-Net and 0.58 for the U-Net RI-GANs, showing that the uncertainties are representative of the actual errors.

The models are also evaluated on an out-of-distribution image from the 30 Doradus region which has a larger dynamic range ($\sim 600$). 
In this example we see that the GU-Net RI-GAN provides significantly better reconstructions and is better at reconstructing the larger dynamic range than the U-Net variant. 
Additionally, the GU-Net provides lower estimated uncertainties. 
Including the measurement information explicitly in the generator clearly also improves the ability of the model to generalise to out-of-distribution data. 
The GU-Net RI-GAN also produces higher quality reconstructions than the CLEAN algorithm, which is used as a baseline for comparison.
We expect that additional improvements in the reconstruction quality and uncertainty estimates can be achieved by training on more representative data, such as images with a larger dynamic range, and by including bright point sources in the training data.

In conclusion, we presented two generative models built on our RI-GAN framework for interferometric image reconstruction. Overall the GU-Net RI-GAN, which includes the measurement operator explicitly in the generator, provides better reconstructions and lower uncertainties than the U-Net variant and generalises better to out-of-distribution data. Using this approach we can provide fast and high-quality image reconstructions of interferometric measurements, while also providing uncertainty estimates that correlate well with the actual errors of the reconstructions.

Further work is needed to perform a more in depth analysis of the statistical significance of the uncertainties to verify whether the generated samples follow the true posterior distribution. One method to do this would be to use estimate the coverage probabilities using the generated samples \citep{hermansCrisisSimulationBasedInference2022,lemosSamplingBasedAccuracyTesting2023}. 

Additionally, more work is needed to evaluate and adapt these models to real radio observations. 
This includes the inclusion of non-coplanar baselines in simulation, addition of bright point sources in the training data, application to larger image sizes and evaluation on images that have a dynamic range more representative of true radio observations.

\section*{Acknowledgements}
We thank Tariq Blecher for help generating MeerKAT visibility coverages.
The authors would like to acknowledge the use of OpenAI's ChatGPT for providing stylistic suggestions but take responsibility for the accuracy of this manuscript.
Matthijs Mars is supported by the UCL Centre for Doctoral Training in Data Intensive Science (STFC Training grant ST/P006736/1).
This work is also partially supported by EPSRC and STFC (grant numbers EP/T517793/1; EP/W007673/1; ST/W001136/1).
This work also used computing equipment funded by the Research Capital Investment Fund (RCIF) provided by UKRI, and partially funded by the UCL Cosmoparticle Initiative. 

\section*{Data Availability}
The code used to generate the results in this paper can be found at \url{https://github.com/astro-informatics/rcGAN}. The trained final models and all necessary data to run them are stored at \url{https://zenodo.org/records/16529320}. The data used in this paper is available upon reasonable request to the corresponding author.



\bibliographystyle{mnras}
\bibliography{Bibliography/lib.bib} 








\bsp	
\label{lastpage}
\end{document}